\documentclass[aps,twocolumn,superscriptaddress,showpacs,amsmath,amssymb,nofootinbib]{revtex4-1}
\pdfoutput=1
\usepackage{graphicx,epsf,amsmath}
\usepackage[]{latexsym}
\usepackage{color}

\newcommand{\be}{\begin{eqnarray}}
\newcommand{\ee}{\end{eqnarray}}

\newcommand{\ket}[1]{\mbox{$\mid #1\,\rangle$}}

\newcommand{\pro}[2]{\mbox{$\langle\, #1 \mid #2\,\rangle$}}
\newcommand{\expec}[1]{\mbox{$\langle\, #1\,\rangle$}}

\renewcommand{\d}{\mbox{${\rm d}$}}
\newcommand{\lp}{\ell_{\rm p}}
\newcommand{\mpl}{m_{\rm p}}
\newcommand{\gn}{G_{\rm N}}
\newcommand{\rh}{R_{\rm H}}
\newcommand{\Rh}{R_{\rm H}}
\begin{document}
%
%
\title{Inner Horizon of the Quantum Reissner-Nordstr\"om Black Holes}
\author{Roberto~Casadio}
\email{casadio@bo.infn.it}
\affiliation{Dipartimento di Fisica e Astronomia,
Alma Mater Universit\`a di Bologna,
via~Irnerio~46, 40126~Bologna, Italy}
\affiliation{I.N.F.N., Sezione di Bologna, viale Berti~Pichat~6/2, 40127~Bologna, Italy}
\author{Octavian~Micu}
\email{octavian.micu@spacescience.ro}
\affiliation{Institute of Space Science, Bucharest,
P.O.~Box MG-23, RO-077125 Bucharest-Magurele, Romania}
\author{Dejan Stojkovic}
\email{ds77@buffalo.edu}
\affiliation{HEPCOS, Department of Physics, SUNY at Buffalo, Buffalo, NY 14260-1500}
\begin{abstract}
We study the nature of the inner Cauchy horizon of a Reissner-Nordstr\"om black
hole in a quantum context by means of the horizon wave-function obtained from
modelling the electrically charged source as a Gaussian wave-function.
Our main finding is that there are significant ranges for the black hole mass
(around the Planck scale) and specific charge for which the probability of realising
the inner horizon is negligible.
This result suggests that any semiclassical instability one expects near the inner
horizon may not occur in quantum black holes.
\end{abstract}
\maketitle
\section{Introduction}
Current literature contains many attempts at quantizing black hole (BH) metrics,
which focus on the purely gravitational degrees of freedom, and yield
a description of the horizon unrelated to the matter state that sources the
geometry~\cite{Greenwood:2008ht}.  
However, this point of view might miss important features emerging from the
highly non-linear nature of the gravitational interaction.
One can think in analogy to the hydrogen atom, and note that its energy
levels, which have no classical counterpart, cannot be simply obtained
by quantising the free electromagnetic field. 
A different perspective is taken in the Horizon Wave Function (HWF)
formalism~\cite{Casadio,C14}, which is instead based on the quantum version of
the Einstein equation relating the size of the gravitational radius (which
can be a horizon) to the (quantum) state of matter. 
This formalism has been applied to several case studies~\cite{Ctest,BEC_BH},
yielding sensible results in agreement with (semi)classical expectations,
and there is therefore hope that it will help our understanding of the quantum nature
of BHs.
\par
In practical terms, the construction of the HWF starts from the spectral decomposition
of the  quantum mechanical state that represents a matter source localized in space.
By expressing the energy in terms of the gravitational (Schwarzschild) radius,
as it would be classically determined according to the Einstein equations,
the spectral decomposition then directly yields the HWF.
The normalised HWF supplies the probability for an observer to detect a gravitational
radius of a certain areal radius, centred around the source in the quantum state that
was used in the first place.
The gravitational radius can then be interpreted as a horizon if the probability of finding
the particle inside it, is reasonably high. 
In other words, the horizon size is necessarily ``fuzzy'' in this QM description,
just like the position of the particles that sources the geometry.
\par
One would expect that quantum modifications of the sort mentioned above
should be important only for small BHs, with masses close to the Planck scale,
and that the quantum corrections to the classical solutions of large mass
BHs should be negligible.
A large BH can have arbitrary small curvature near its horizon,
and we have numerous tests of gravity in such low density regimes.
However, BHs are very peculiar objects, in that they trap any signal inside
the horizon {\em by definition\/}, no matter how weak tidal forces are in its
neighbourhood, and it may happen that they are better described as macroscopic
quantum states (see for instance, the Bose-Einstein Condensate model of
Refs.~\cite{dvali}). 
\par
Charged BHs were subject to many theoretical studies in the
past~\cite{Wang:2009ay}.
In this work, we shall extend the HWF formalism to the Reissner-Nordstr\"om (RN)
geometry and investigate the probability for an electrically charged source
represented by a Gaussian wave-packet to be a BH, and to have an actual 
inner horizon.
In the semiclassical approach, the latter is a Cauchy horizon and is
sometimes associated with an instability known as ``mass inflation'':
any small matter perturbation will blue-shift unboundedly just outside this
horizon, and inevitably produce a large deformation to the background
geometry~\cite{poisson}.
Although the existence of this effect is still debated
(see, e.g.~Refs.~\cite{Dokuchaev:2013uda}),
it is clear from the classical causal structure of the RN geometry that, if there
is matter falling through the outer horizon, it should accumulate outside the
inner Cauchy horizon, and eventually lead to a large backreaction there.
On the other hand, matter inside the inner horizon could escape the
Cauchy horizon and produce a deformation as well.
It is thus interesting to study under which conditions the inner horizon 
survives in the QM treatment.
\section{The HWF Formalism}
\label{RNsystem}
The formalism introduced in Refs.~\cite{Casadio,C14}
is based on lifting the gravitational radius $\rh$ of a QM system
to the rank of a quantum operator.
The coordinate $r$ in a spherical metric is invariantly related to the geometrical
area $4\,\pi\,r^2$ of the sphere centred on the origin $r=0$, therefore being
a suitable candidate for an observable in the quantum theory.
Moreover, the horizon radius $r=\Rh$ represents the location of trapping
surfaces (surfaces where the escape velocity equals the speed of light)
and thus determines the causal structure of the space-time, which again
is likely an observable property in the quantum theory. 
\par
We remind the readers that in a neutral spherically symmetric system,
\be
\Rh(r)
=
2\,\lp\,\frac{M(r)}{\mpl}
\ ,
\label{RH}
\ee
where $\lp$ is the Planck length and $\mpl$ the Planck mass~\footnote{We shall use
units with $c=k_B=1$, and always display the Newton constant $\gn=\lp/\mpl$,
so that $\hbar=\lp\,\mpl$.}, and 
\be
M(r,t)
=
4\,\pi\int_0^r \rho(\bar r,t)\,\bar r^2\,\d \bar r
\label{M}
\ee
is the Misner-Sharp mass.
Here $M=M(r,t)$ represents the {\em total energy\/} inside the sphere of 
area $4\,\pi\,r^2$ (thus, roughly speaking, including the negative gravitational energy)
and is related to the energy density $\rho$ of the source via the {\em flat\/} space
volume.
A specific value of $r$ is a trapping surface if $\Rh(r)=r$, while, if $\Rh(r)<r$,
the gravitational radius is still well-defined but does not correspond to any causal
surface.
\par
Going back to the HWF formalism, let us start from QM states
representing {\em spherically symmetric\/} objects, which are
{\em localized in space\/} and {\em at rest\/} in the chosen reference frame.
Such particles are consequently described by wave-functions  
$\psi_{\rm S}\in L^2(R^3)$,
which we assume can be decomposed into energy eigenstates,
\be
\ket{\psi_{\rm S}}
=
\sum_E\,C(E)\,\ket{\psi_E}
\ ,
\ee
where the sum represents the spectral decomposition in Hamiltonian
eigenmodes,
\be
\hat H\,\ket{\psi_E}=E\,\ket{\psi_E}
\ .
\ee
The actual Hamiltonian $H$ needs not be specified yet~\footnote{This is where,
for instance, the self-gravity of the particle may enter.}.
\par
The expression of the Schwarzschild radius in Eq.~\eqref{RH} can be inverted
to obtain
\be
M
=
\mpl\,\frac{\rh}{2\,\lp}
\ ,
\ee
and remembering that $M$ represents the total energy, it can be used to define the (unnormalized) ``horizon wave-function'' as
\be
\tilde\psi_{\rm H}(\rh)
=
C\left(\mpl\,{\rh}/{2\,\lp}\right)
\ .
\ee
\par 
The normalisation of $\tilde\psi_{\rm H}$ is finally fixed by means of the scalar
product
\be
\pro{\psi_{\rm H}}{\phi_{\rm H}}
=
4\,\pi\,\int_0^\infty
\psi_{\rm H}^*(\rh)\,\phi_{\rm H}(\rh)\,\rh^2\,\d \rh
\ ,
\label{normH}
\ee
again in agreement with the geometrical meaning of the variable $\rh$ 
as yielding the area $4\,\pi\,\rh^2$ of the corresponding sphere.
Our interpretation of the normalised $\psi_{\rm H}$ is then that it yields the probability
for an observer to detect a horizon (necessarily ``fuzzy'', like the position of the
particle itself) of areal radius $r=\rh$, associated with the particle in the quantum
state $\psi_{\rm S}$.
\par
In more details, starting from the he wave-function $\psi_{\rm H}$ associated
with $\psi_{\rm S}$, we can now calculate the probability density for the particle
to lie inside its own horizon of radius $r=\rh$:
\be
{\mathcal P}_<(r<\rh)
=
P_{\rm S}(r<\rh)\,{\mathcal P}_{\rm H}(\rh)
\ ,
\label{PrlessH}
\ee
where
\be
P_{\rm S}(r<\rh)
=
4\,\pi\,\int_0^{\rh}
|\psi_{\rm S}(r)|^2\,r^2\,\d r
\ee
is the probability that the particle is inside a sphere of radius $r=\rh$,
and
\be
{\mathcal P}_{\rm H}(\rh)
=
4\,\pi\,\rh^2\,|\psi_{\rm H}(\rh)|^2
\label{Ph}
\ee
is the probability density that the sphere of radius $r=\rh$ is a horizon.
Finally, the probability that the particle is a
black hole will be obtained by integrating~\eqref{PrlessH} over all possible
values of the horizon radius, namely
\be
P_{\rm BH}
=
\int_0^\infty 
{\mathcal P}_<(r<\rh)\,\d \rh
\ .
\label{PBH}
\ee
\par
To anticipate the calculations in the next section, we note that electrically
charged BHs have two gravitational radii, and we will therefore define two
corresponding operators, namely $\hat R_\pm$.
We shall then define a HWF for each one, and obtain the probabilities
for both the inner and outer horizon to exist.
\section{Electrically charged spherical sources}
\label{RNsystem}
We start from the RN metric, which can be written as
\be
\d s^2
=
-f\,\d t^2
+f^{-1}\,\d r^2
+r^2
\left(d\theta^2 + \sin^2 \theta\,d\phi^2\right)
\ ,
\label{RN}
\ee
with 
\be
f=
1- \frac{2\, \lp\, M}{\mpl \,r}+\frac{Q^2}{r^2}
\ ,
\ee
where $M$ and $Q$ respectively represent the ADM mass and charge of the source.
It is now convenient to introduce the specific charge
\be
\alpha
=
\frac{|Q|\, \mpl}{\lp\, M}
\ .
\label{alpha}
\ee
The case $\alpha=0$ reduces to the neutral Schwarzschild metric,
and shall not be reconsidered here (see Refs.~\cite{Casadio,C14,Ctest,BEC_BH}).
For $0<\alpha<1$, the above metric contains two horizons, namely 
\be
R_{\pm}
&=&
\lp\, \frac{M}{\mpl}\pm\sqrt{\left(\lp\, \frac{M}{\mpl}\right)^2 - Q^2}
\nonumber
\\
&=&
\lp\, \frac{M}{\mpl}\left(1\pm\sqrt{1- \alpha^2}\right)
\ ,
\label{R+-}
\ee
and represents a BH.
The two horizons overlap for $\alpha=1$, the so-called extremal BH case,
while for $\alpha>1$ no horizon exists and the
central singularity is therefore accessible to an outer observer.
This is the prototype of a naked singularity, but we shall not consider this case
here (the corresponding HWF is the topic of Ref.~\cite{Casadio:2015sda}).
\par
We shall now investigate the case $0<\alpha \le 1$, which classically possesses
at least one horizon, from a QM perspective.
We first determine the HWFs and then calculate the probabilities for both
the inner and outer horizon to exist.
For this purpose, the classical relations~\eqref{R+-} will be lifted to
the rank of equations for the operators $\hat R_\pm$ and
$\hat M$, which are chosen to act multiplicatively on the HWF
(with the specific charge $\alpha$ viewed as a simple parameter).
\subsection{HWF for Gaussian source}
\label{Gparticle}
The source for the RN space-time is taken to be an electrically
charged massive particle at rest in the origin of the reference frame,
represented by the spherically symmetric Gaussian wave-function
\be
\psi_{\rm S}(r)
=
\frac{e^{-\frac{r^2}{2\,\ell^2}}}{\ell^{3/2}\,\pi^{3/4}}
\ .
\label{psis}
\ee
The width of the Gaussian $\ell$ is assumed to be the minimum 
compatible with the Heisenberg uncertainty principle, that is
\be
\ell
=
\lambda_m
\simeq
\lp\,\frac{\mpl}{m}
\ ,
\ee
where $\lambda_m$ is the Compton length of the particle of rest mass $m$
(this assumptions is thoroughly investigated in Ref.~\cite{C14}).
The spectral decomposition of Eq.~\eqref{psis} is easily obtained from
assuming the relativistic mass-shell relation in flat space,
\be
M^2=p^2+m^2
\ ,
\label{mass shell}
\ee
and by going to momentum space,
\be
\psi_{\rm S}(p)
=
\frac{e^{-\frac{p^2}{2\,\Delta^2}}}{\Delta^{3/2}\,\pi^{3/4}}\, \label{psi_p}
\ ,
\label{psip}
\ee
where $p^2=\vec p\cdot\vec p$ is the square modulus of the spatial momentum,
and the width $
\Delta
=
\mpl\,{\lp}/{\ell}
\simeq
m$.
\par
For $\alpha<1$, one can now write a HWF for each of the two horizons in Eq.~\eqref{R+-}.
In fact, the total energy $M$ can be expressed in terms of the horizon radii as
\be
\lp\,\frac{\hat M}{\mpl}
=
\frac{\hat R_{+}+\hat R_{-}}{2}
\ ,
\label{EofRpm}
\ee
and
\be
\hat R_{\pm}
=
\hat R_{\mp}\,
\frac{1\pm\sqrt{1-\alpha^2}}{1\mp\sqrt{1-\alpha^2}}
\ ,
\label{R-R+}
\ee
where $M$, $R_{+}$, and $R_{-}$ were promoted to operators
$\hat{M}$, $\hat R_{+}$, and $\hat R_{-}$
related to the corresponding observables.
It is important to remark that our choice is not unique, because of
the usual ambiguities that emerge when going from a classical to
the quantum formalism.  
The unnormalized HWFs for $R_{+}$ and $R_{-}$ are obtained 
by expressing $p$ from Eq.~\eqref{mass shell} in terms of $M$
in Eq.~\eqref{EofRpm}, and then replacing one of the relations in
Eq.~\eqref{R-R+} into Eq.~\eqref{psi_p}.
These manipulations yield
\be
\psi_{\rm H}(R_{\pm})
&=&
\mathcal{N_\pm}\,
\Theta\left(R_{\pm}-R_{\rm min \pm}\right)
\nonumber
\\
&&
\times\, 
\exp\left\{-\frac{\mpl^2\,R_{\pm}^2}{2\,\Delta^2\,\lp^2\,(1\pm\sqrt{1-\alpha^2})^2}\right\}
\ .
\label{psih}
\ee
The step function in the first line above accounts for the minimum energy $M=m$
in the spectral decomposition of the wave-function~\eqref{psis}, which corresponds to
\be
R_{\rm min \pm}
=
\lp\,\frac{m}{\mpl}\left(1\pm\sqrt{1 - \alpha^2}\right)
\ .
\label{Rminalpha}
\ee
Finally, the normalisations $\mathcal{N_\pm}$ are fixed again by using the scalar product
in Eq.~\eqref{normH} for both $R_\pm$~\footnote{Explicit
expressions of $\mathcal{N_\pm}$ are very cumbersome and not particularly significant,
thus shall be omitted throughout the paper.}.
\par
The probability density that the particle lies inside its own horizons of size
$r=R_{\pm}$ can now be calculated starting from the wave-functions~\eqref{psih}
associated with~\eqref{psis} as
\be
{\mathcal P}_{<\pm}(r<R_{\pm})
=
P_{\rm S}(r<R_{\pm})\,{\mathcal P}_{\rm H}(R_{\pm})
\ ,
\label{PrlessH}
\ee
where
\be
P_{\rm S}(r<R_{\pm})
=
4\,\pi\,\int_0^{R_{\pm}}
|\psi_{\rm S}(r)|^2\,r^2\,\d r
\ee
is the probability that the particle is inside the sphere $r=R_{\pm}$,
and
\be
{\mathcal P}_{\rm H}(R_{\pm})
=
4\,\pi\,R_{\pm}^2\,|\psi_{\rm H}(R_{\pm})|^2
\label{Ph}
\ee
is the probability density that the sphere $r=R_{\pm}$ is a horizon.
Finally, one can integrate~\eqref{PrlessH} over all possible
values of the horizon radius $R_+$ to find the probability for the particle
described by the wave-function~\eqref{psis} to be a BH, namely
\be
P_{\rm BH+}
=
\int_{R_{\rm min+}}^\infty {\mathcal P}_{<+}(r<R_+)\,\d R_+
\ .
\label{PBH+}
\ee
The analogous quantity for $R_-$,
\be
P_{\rm BH-}
=
\int_{R_{\rm min-}}^\infty {\mathcal P}_{<-}(r<R_-)\,\d R_-
\ ,
\label{PBH-}
\ee
will instead be viewed as the probability that the particle lies further inside 
its inner horizon.
It is already clear from these definitions that $P_{\rm BH-}<P_{\rm BH+}$,
and it is only when $P_{\rm BH-}$ is significantly close to one that we can 
say that both $R_-$ and $R_+$ are physically realised.
\subsection{Inner and outer horizon probabilities}
\label{results}
The probabilities $P_{\rm BH\pm}$ can only be computed numerically,
and we shall therefore display their behaviour graphically.
\par
\begin{figure}[t]
\centering
\raisebox{4.5cm}{${\mathcal P}_{<+}$}
\includegraphics[scale=0.92]{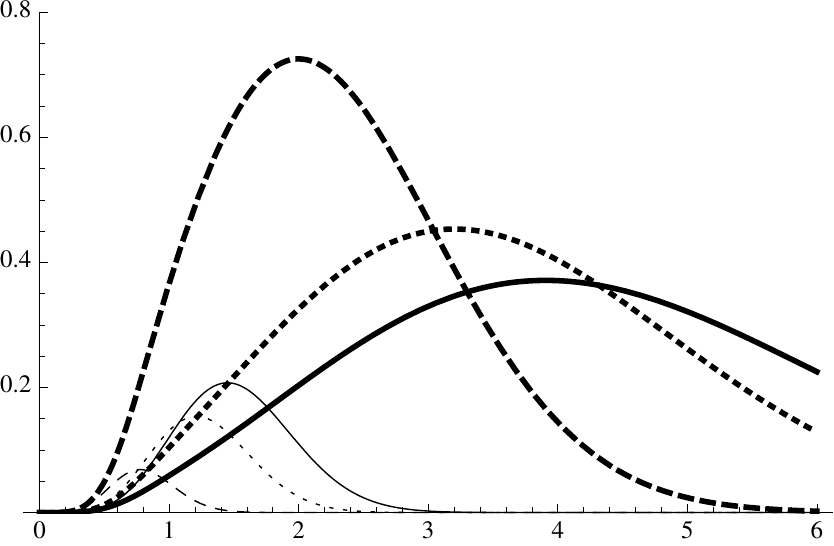}
\\
\hspace{7cm}$r/\lp$
\\
\raisebox{4.5cm}{${\mathcal P}_{<-}$}
\includegraphics[scale=0.92]{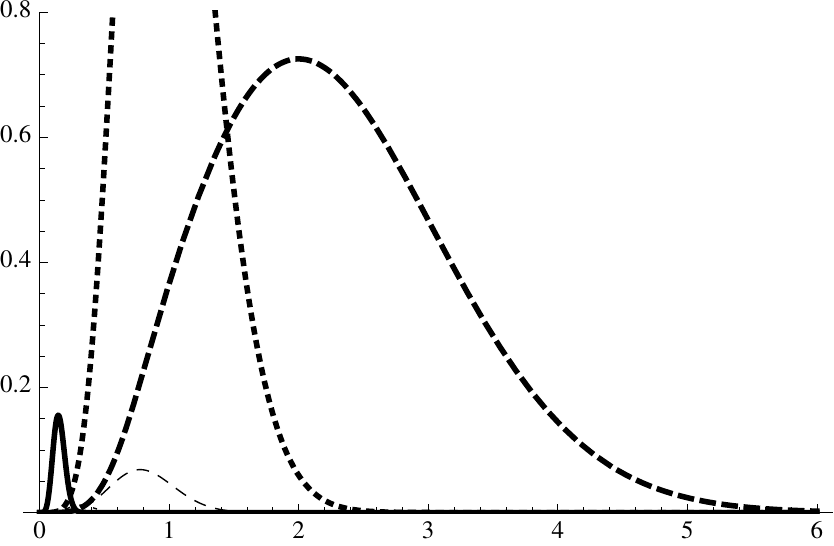}
\\
\hspace{7cm}$r/\lp$
\caption{Top panel: probability density ${\mathcal P}_{<+}$ in Eq.~\eqref{PrlessH}
that the particle is inside its outer horizon $r=R_{+}$,
for $m=2\,\mpl$ (thick lines) and $m=0.5\,\mpl$ (thin lines)
with $\alpha=0.3$ (continuous lines), $\alpha=0.8$ (dotted lines) and $\alpha=1$
(dashed lines).
Botton panel: probability density ${\mathcal P}_{<-}$ in Eq.~\eqref{PrlessH}
that particle is inside its inner horizon $r=R_{-}$,
for $m=2\,\mpl$ (thick lines) and $m=0.5\,\mpl$ (thin lines)
with $\alpha=0.3$ (continuous lines), $\alpha=0.8$ (dotted lines) and $\alpha=1$
(dashed lines). 
For $\alpha=1$, the two horizons coincide and ${\mathcal P}_{<-}={\mathcal P}_{<+}$.
\label{prob<}}
\end{figure}
In the upper panel of Fig.~\ref{prob<} we show the probability density that
the particle lies inside the outer horizon $r=R_{+}$ from Eq.~\eqref{PrlessH}
for two values of the Gaussian width $\ell=\lambda_m\sim m^{-1}$
(above and below the Planck scale).
This probability clearly decreases when $m$ decreases
below the Planck mass, which corresponds to $\ell$ increasing above $\lp$.
A similar analysis is presented in the lower panel of the same figure for the
probability density that the particle lies inside the inner horizon $r=R_{-}$.
It is obvious that the probabilities in the second case are much smaller with
decreasing $\alpha$.
As expected, the two probability densities are identical for the extremal
case $\alpha=1$ (thick and thin dashed lines), 
since the two horizons coincide.   
\begin{figure}[t]
\centering
\raisebox{4.5cm}{$P_{\rm BH}$}
\includegraphics[scale=0.92]{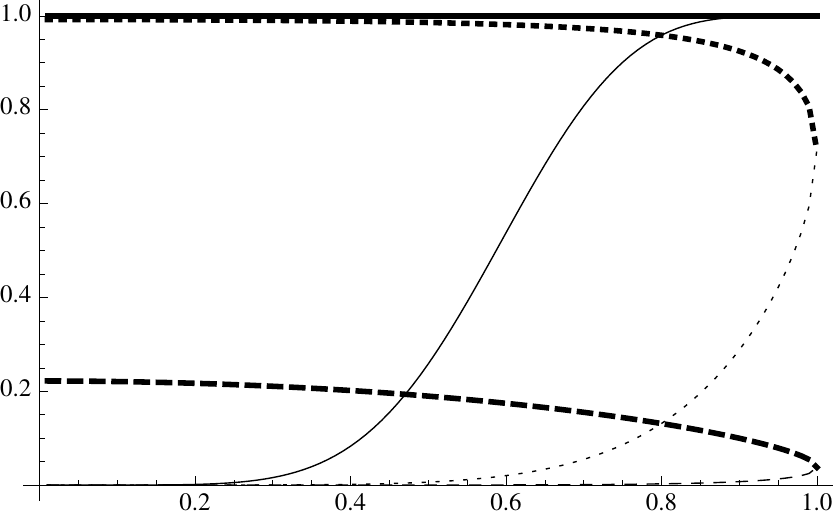}
\\
\hspace{7cm}$\alpha$
\caption{Probability $P_{\rm BH+}$ in Eq.~\eqref{PBH+} for the particle to be a BH (thick lines)
and $P_{\rm BH-}$ in Eq.~\eqref{PBH-} for the particle to be inside its inner horizon (thin lines)
as functions of $\alpha$ for $m=2\,\mpl$ (continuous line), $m=\mpl$ (dotted line)
and $m=0.5\,\mpl$ (dashed line).
For $\alpha=1$ the two probabilities merge. 
\label{PBHplot1}}
\end{figure}
\par
The probabilities $P_{\rm BH\pm}$ are obtained by performing the
integrations~\eqref{PBH+} and \eqref{PBH-}.
The plot in Fig.~\ref{PBHplot1} shows these probabilities as
functions of $\alpha$ for values of the particle mass above, equal to and
below the Planck mass.
When analyzing the outer horizon, one notices that $P_{\rm BH+}$ stays very
close to one for mass values larger than the Planck scale.
However, for $m\lesssim \mpl$ (when the width of the Gaussian wave-packet
$\ell\gtrsim \lp$), $P_{\rm BH+}$ clearly decreases as the BH specific charge
increases to one.
Note that this probability is not exactly zero even for values of the mass smaller
than $\mpl$.
For instance, in the case $m=0.5\, \mpl$, corresponding to a width $\ell=2\,\lp$
of the Gaussian wave-packet, $P_{\rm BH+}\simeq 0.2$ for a considerable range
of values of $\alpha$.
It only decreases below $0.1$ when $\alpha$ approaches one, therefore
when the BH becomes maximally charged.
The situation is very different for the inner horizon.
The same plot shows that the probability $P_{\rm BH-}$ starts from almost
zero for small values of the charge-to-mass ratio and increases with $\alpha$.
The larger the mass of the particle, the smaller the value of $\alpha$ for
which the probability starts to become significant.
Still, there is a considerable range of values of the specific charge for which,
while $P_{\rm BH+}\simeq 1$ thus making the object a BH, 
the probability for the inner horizon to exist is approximately zero. 
\par
\begin{figure}[t]
\centering
\raisebox{4.5cm}{$P_{\rm BH}$}
\includegraphics[scale=0.92]{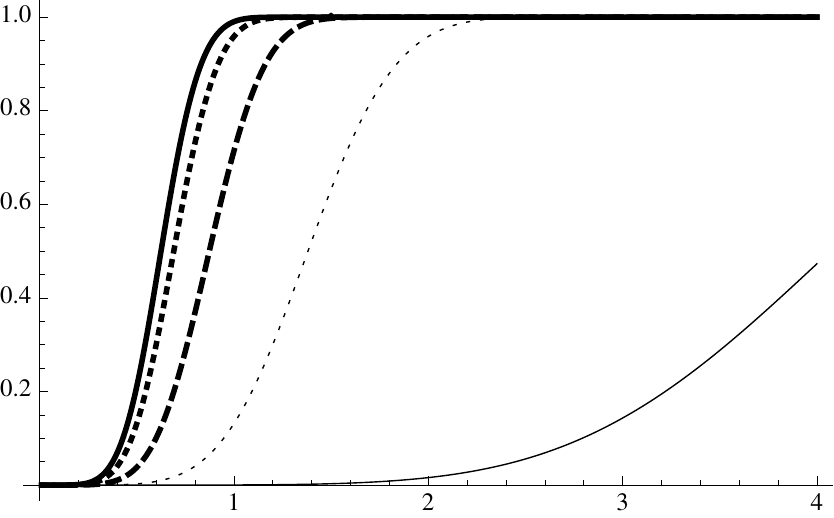}
\\
\hspace{7cm}$m/\mpl$
\caption{Probability $P_{\rm BH+}$ in Eq.~\eqref{PBH+} for the particle to be a BH (thick lines)
and $P_{\rm BH-}$ in Eq.~\eqref{PBH-} for the particle to be inside its inner horizon (thin lines)
as functions of the mass for $\alpha=0.3$ (continuous line), $\alpha=0.8$ (dotted line) and
$\alpha=1$ (dashed line).
For $\alpha=1$ thick and thin dashed lines overlap. 
\label{PBHplot2}}
\end{figure}
Fig.~\ref{PBHplot2} shows the probabilities $P_{\rm BH\pm}$
as functions of the mass $m$ for $\alpha=0.3$, $0.8$ and $1$.
From this plot it becomes clear that for smaller values of $\alpha$,
the probability $P_{\rm BH+}$ starts to increase from zero to one at smaller
values of $m$.
The opposite is true when analyzing $P_{\rm BH-}$.
For the smallest value of the charge-to-mass ratio considered here,
$\alpha=0.3$, it is only around a particle mass of $m\simeq 6\,\mpl$
that both probabilities $P_{\rm BH+}$ and $P_{\rm BH-}$ have values close to one,
while $P_{\rm BH+}$ already increases to one around $\mpl$.
This means that in the range of masses between $\mpl$ and $6\,\mpl$,
the probability $P_{\rm BH+}\simeq 1$ while $P_{\rm BH-}\simeq 0$.
This mass range increases even more for smaller values of the specific charge, 
but it decreases to zero in the maximally charged limit.
\par
Our main finding is therefore that there exists a considerable parameter
space for $m$ (around the Planck scale) and $\alpha< 1$ in which
\be
P_{\rm BH+}\simeq 1
\quad
{\rm and}
\quad
P_{\rm BH-}\simeq 0
\ .
\label{P10}
\ee
Whether the particle is a BH or not is dictated by the existence of
the outer horizon, therefore we interpret $P_{\rm BH+}\simeq 1$ as
meaning that the particle is (most likely) a BH.
However, the presence of an inner horizon at $r=R_{-}$
is important in light of the ``mass inflation'' instability and peculiar
features of Cauchy horizons.
Eq.~\eqref{P10} therefore means that the particle is (most likely) a BH,
but no such peculiarities are expected to occur.

\subsection{Generalised uncertainty principle}
\label{GUP}
The uncertainty in the horizon size for a neutral BH
was already investigated previously.
In particular, it was shown to grow linearly with the BH mass
and lead to a generalised uncertainty principle (GUP) in Ref.~\cite{Casadio}.
We can here repeat the same arguments for the outer
horizon of the charged BH, and obtain similar results.
\par
We first note that the expectation value
\be
\expec{\hat R_{+}}
&=&
4\,\pi\,\int_{R_{\rm min+}}^{\infty}
|\psi_{\rm H}(R_{+})|^2\,R_{+}^3\,\d R_{+}
\nonumber
\\
&=&
\frac{4\left(1\,+ \sqrt{1-\alpha^2}\right)}{2\,+e\,\sqrt{\pi}\,{\rm erfc}{(1)}}\,
\frac{\lp^2}{\ell}\label{dR+}
\nonumber
\\
&=&
R_+(\bar M)
\ ,
\ee
reproduces exactly the classical expression of $R_+$ in
Eq.~\eqref{R+-} for $\ell=\lambda_m\sim m^{-1}$ and
$\bar M=4\,m/[2\,+e\,\sqrt{\pi}\,{\rm erfc}{(1)}]\simeq 1.45\,m$~\footnote{This mass
renormalisation, with $\bar M>m$, can be easily understood by noting that the source
wave-function $\psi_{\rm S}$ contains energy contributions from momenta $p>0$.}.
From
\be
\expec{\hat R_{+}^2}
&=&
4\,\pi\,\int_{R_{\rm min+}}^{\infty}
|\psi_{\rm H}(R_{+})|^2\,R_{+}^4\,\d R_{+}
\nonumber
\\
&=&
\frac{\left(1\,+ \sqrt{1-\alpha^2}\right)^2 \left(10\,+3\,e\,\sqrt{\pi}\,{\rm erfc}(1)\right)}
{2\left(2\,+e\,\sqrt{\pi}\,{\rm erfc}(1)\right)}\,
\frac{\lp^4}{\ell^2}
\nonumber
\\
&\simeq&
R_+^2(\bar M)
\ ,
\quad
\label{dR2+}
\ee
one can then calculate the uncertainty 
\be
\Delta R_{+}
=
\sqrt{\expec{\hat R_{+}^2} - \expec{\hat R_{+}}^2}
\simeq
R_+
\sim
m
\ ,
\label{DRH+}
\ee
which, like in the neutral Schwarzschild case, grows linearly with the mass $m$
of the source.
This signals the fact that the state of such objects would remain QM even in an
astrophysical regime, where we instead expect the horizon has a sharp location,
and supports alternative models of large BHs, such as the ones in Refs.~\cite{dvali}.
\par
\begin{figure}[t]
\centering
\raisebox{3.5cm}{$\frac{\Delta r}{\lp}$}
\includegraphics[width=8cm]{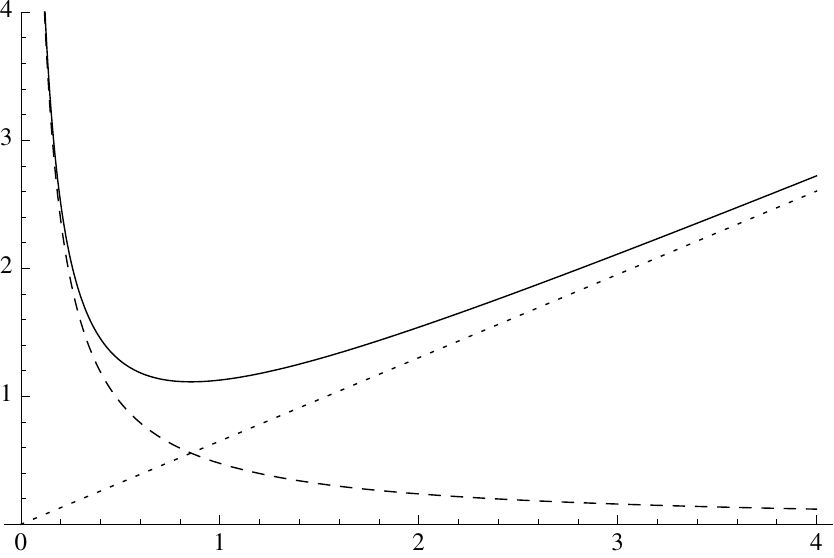}
\\
\hspace{6cm}$\Delta p/\mpl$
\caption{Uncertainty relation~\eqref{effGUP} (solid line) as a combination
of the QM uncertainty (dashed line) and the uncertainty
in horizon radius (dotted line), for $\gamma=1$.
\label{pGUP}}
\end{figure}

If we now combine the horizon uncertainty~\eqref{DRH+} with the usual QM
uncertainty in the radial size of the source,
\be
\Delta r^2
&=&
4\,\pi\!\!
\int_0^{\infty}
\!\!
|\psi_{\rm S}(r)|^2\,r^4\,\d r
-
\left(\!
4\,\pi\!\!
\int_0^{\infty}
\!\!
|\psi_{\rm S}(r)|^2\,r^3\,\d r\!
\right)^2
\nonumber
\\
&\simeq&
\ell^2
\ ,
\label{Dr}
\ee
we can finally obtain a total uncertainty
\be
\Delta r
&\equiv&
\sqrt{\expec{\Delta r^2}}
+
\gamma\,
\sqrt{\expec{\Delta R_{+}^2}}
\nonumber
\\
&\simeq&
\lp\,\frac{\mpl}{\Delta p}
+
\gamma\,\lp\,\frac{\Delta p}{\mpl}
\ ,
\label{effGUP}
\ee
where $\gamma$ is a coefficient of order one.
The result is plotted in Fig.~\ref{pGUP}, where it is also compared
to the usual Heisenberg uncertainty in the size $\Delta r$ of a state
with an uncertainty in momentum given by
\be
\Delta p^2
&=&
4\,\pi\!\!
\int_0^{\infty}
\!\!
|\psi_{\rm S}(p)|^2\,p^4\,\d p
-
\left(\!
4\,\pi\!\!
\int_0^{\infty}
\!\!
|\psi_{\rm S}(p)|^2\,p^3\,\d p\!
\right)^2
\nonumber
\\
&\simeq&
\Delta^2
\simeq
\ell^{-2}
\ .
\label{Dp}
\ee
We can therefore conclude that the outer horizon behaves qualitatively like
the neutral Schwarzschild radius.
\section{Conclusions and outlook}
In this work we transparently applied some basic QM methods to the case
of the RN BH.
In our formalism, the location of the horizon is not given by a sharp classical
value, instead it is described by a quantum wave-function with associated
uncertainties.
In addition, one can define a quantity which corresponds to the probability
for a horizon to be formed.
Since the RN BH has the inner and outer horizon, the structure of a quantum
BH might be quite different from the structure of the classical one. 
\par
We first note here that, if one models a BH as a {\em point-like\/}
source (i.e.~one very narrow Gaussian wave-function), the uncertainty
in the location of the outer horizon $\Delta R_{+}$ grows with the mass
of the source (since $\ell \sim m^{-1}$).
This is in agreement with many postulated GUPs in the presence of
gravity~\cite{Hossenfelder:2012jw,Chang:2011jj},
but implies that when the source is of astrophysical size, the fluctuations
become unacceptably large.
This might imply that there must be a regime where the GUP gets replaced
by the standard (non-gravitational) uncertainty principle of QM.
In other words, the parameter $\gamma$ in Eq.~(\ref{effGUP}) becomes
very small in the semi-classical regime. 
\par
The other (somewhat related) option is that the single Gaussian
source~\eqref{psis} does not appear to be a sensible model for large BHs.
The fact that the uncertainty in the horizon location does not decrease for increasing
rest mass $m$, might imply that a semiclassical behavior cannot be recovered
at all in such a setup~\cite{Casadio}.
This result was remarked in Ref.~\cite{BEC_BH}, where it was then shown
that modelling a BH as a large number of light constituents~\cite{dvali} does
not suffer of this limitation.
Because of that, we cannot here extend our findings to arbitrarily large
BH mass straightforwardly.
\par
We also note that modelling the source with some shape different
from the Gaussian will not make qualitative differences.
For example, using a step function will give a result in qualitative
agreement with the Gaussian of similar width.
In fact, one can reliably approximate any localised state $\psi_{\rm S}$
with a superposition of Gaussians.
Given the linearity of the formalism, a superposition of, say, $N$
Gaussians of roughly similar mass $m$ will lead to a total HWF given
by a superposition of the corresponding HWFs.
In particular, if the width $\ell$ of the Gaussians are very narrow
($\ell \leq  \lp$), we can understand what happens by simply replacing
the superposition with one Gaussian of same $\ell$ and mass equal to $N\,m$.
The latter will increase $\expec{\hat R_H} \gg \ell$, and also increase
the probability $P_{\rm BH}$.
However, if $\ell \sim 1/ N\,m$, the uncertainty $\Delta R_H\sim \expec{\hat R_H}$
and the system will never look classical. 
In contrast, if we model the source as $N$ Gaussians of large width $\ell \gg \lp$,
but $m\ll \mpl$, like in the Bose-Einstein Condensate model of
Dvali and Gomez~\cite{dvali}, then $\expec{\hat R_H}\sim R_H$ and 
$\Delta R_H \ll \expec{\hat R_H}$ for large N, which does instead look classical.
If the source has electrical charge, the above argument then holds for $R_+$.
Since the $N$ constituents of the condensate have width $\ell\sim R_+$,
it immediately follows that the inner horizon $R_-$ will have very small
probability to exist (expect perhaps in a near-extremal configuration).
\par
To conclude, we can speculate that, at least in a quantum regime of BH masses,
say $m\lesssim 10\,\mpl$, quantum fluctuations around the inner horizon are
strong enough to prevent the instability expected according to the semiclassical
analysis.
More generally, the probability that any instability occurs will be as small as the
probability $P_{\rm BH-}$ that the source is located inside the inner horizon.
\subsection*{Acknowledgments}
R.C.~was supported in part by the INFN grant FLAG.
O.M.~was supported in part by research grant UEFISCDI project PN-II-RU-TE-2011-3-0184.
D.S.~was partially supported by the US National Science Foundation,
under Grant No.~PHY-1066278 and PHY-1417317.

\begin{thebibliography}{99}
%
%
\bibitem{Greenwood:2008ht}
  E.~Greenwood and D.~Stojkovic,
 JHEP \ 032P \ 0408;
%
  A.~Saini and D.~Stojkovic,
  Phys.\ Rev.\ D {\bf 89}, no. 4, 044003 (2014);
%
  T.~Vachaspati, D.~Stojkovic and L.~M.~Krauss,
  Phys.\ Rev.\ D {\bf 76}, 024005 (2007);
%
  T.~Vachaspati and D.~Stojkovic,
  Phys.\ Lett.\ B {\bf 663}, 107 (2008).
%
\bibitem{Casadio} 
R.~Casadio,
``Localised particles and fuzzy horizons: A tool for probing Quantum Black Holes,''
arXiv:1305.3195 [gr-qc];
%
R.~Casadio and F.~Scardigli,
Eur.\ Phys.\ J.\ C {\bf 74}, 2685 (2014);
%
\bibitem{C14}
R.~Casadio,
``Horizons and non-local time evolution of quantum mechanical systems,''
arXiv:1411.5848 [gr-qc].
%
\bibitem{Ctest}
R.~Casadio, O.~Micu and F.~Scardigli,
Phys.\ Lett.\ B {\bf 732} (2014) 105
[arXiv:1311.5698 [hep-th]].
%
\bibitem{BEC_BH}
R.~Casadio, A.~Giugno, O.~Micu and A.~Orlandi,
Phys.\ Rev.\ D {\bf 90} (2014) 8,  084040
[arXiv:1405.4192 [hep-th]].
%
\bibitem{dvali}
G.~Dvali and C.~Gomez,
``Black Holes as Critical Point of Quantum Phase Transition'',
arXiv:1207.4059 [hep-th];
Phys.\ Lett.\ B {\bf 719}, 419 (2013);
Phys.\ Lett.\ B {\bf 716}, 240 (2012);
Fortsch.\ Phys.\  {\bf 61}, 742 (2013);
%
R.~Casadio and A.~Orlandi,
JHEP {\bf 1308},  025 (2013).
%
\bibitem{Wang:2009ay} 
  J.~E.~Wang, E.~Greenwood and D.~Stojkovic,
  Phys.\ Rev.\ D {\bf 80}, 124027 (2009)
  [arXiv:0906.3250 [hep-th]].
%
\bibitem{poisson}
  E.~Poisson and W.~Israel,
  Phys.\ Rev.\ D {\bf 41} (1990) 1796.
%
\bibitem{Dokuchaev:2013uda} 
  V.~I.~Dokuchaev,
  Class.\ Quant.\ Grav.\  {\bf 31}, 055009 (2014)
  [arXiv:1309.0224 [gr-qc]].
%
  E.~Brown and R.~B.~Mann,
  Phys.\ Lett.\ B {\bf 694}, 440 (2011)
  [arXiv:1012.4787 [hep-th]].
 %
  E.~G.~Brown, R.~B.~Mann and L.~Modesto,
  Phys.\ Rev.\ D {\bf 84}, 104041 (2011)
  [arXiv:1104.3126 [gr-qc]].
 %
\bibitem{Casadio:2015sda} 
  R.~Casadio, O.~Micu and D.~Stojkovic,
  arXiv:1503.02858 [gr-qc].
%
\bibitem{Hossenfelder:2012jw}
  S.~Hossenfelder,
  Living Rev.\ Rel.\  {\bf 16} (2013) 2
  [arXiv:1203.6191 [gr-qc]].
%
\bibitem{Chang:2011jj}
  L.~N.~Chang, Z.~Lewis, D.~Minic and T.~Takeuchi,
  Adv.\ High Energy Phys.\  {\bf 2011} (2011) 493514
  [arXiv:1106.0068 [hep-th]].
%
\end{thebibliography}
\end{document}